\documentclass[twocolumn]{aastex631}

\usepackage{CJK}

\usepackage{bm}

\shorttitle{Patches of switchbacks}
\shortauthors{Shi et al.}
\graphicspath{{./}{figures/}}

\begin{document}
\begin{CJK*}{UTF8}{gbsn}

\title{Patches of magnetic switchbacks and their origins}

\correspondingauthor{Chen Shi}
\email{cshi1993@ucla.edu}

\author[0000-0002-2582-7085]{Chen Shi(时辰)}
\affiliation{Department of Earth, Planetary, and Space Sciences, University of California, Los Angeles \\
Los Angeles, CA, USA}

\author[0000-0002-4440-7166]{Olga Panasenco}
\affiliation{Advanced Heliophysics Inc. \\
Los Angeles, CA, USA}

\author[0000-0002-2381-3106]{Marco Velli}
\affiliation{Department of Earth, Planetary, and Space Sciences, University of California, Los Angeles \\
Los Angeles, CA, USA}

\author[0000-0003-2880-6084]{Anna Tenerani}
\affiliation{Department of Physics, The University of Texas at Austin, \\
     TX 78712, USA}
     
\author[0000-0003-1138-652X]{Jaye L. Verniero}
\affiliation{Heliophysics Science Division, NASA Goddard Space Flight Center, \\ Greenbelt, MD 20771, USA}

\author[0000-0002-1128-9685]{Nikos Sioulas}
\affiliation{Department of Earth, Planetary, and Space Sciences, University of California, Los Angeles \\
Los Angeles, CA, USA}

\author[0000-0001-9570-5975]{Zesen Huang (黄泽森)}
\affiliation{Department of Earth, Planetary, and Space Sciences, University of California, Los Angeles \\
Los Angeles, CA, USA}

\author[0000-0003-2393-3834]{A. Brosius}
\affiliation{Heliophysics Science Division, NASA Goddard Space Flight Center, \\ Greenbelt, MD 20771, USA}
\affiliation{Department of Atmospheric Science, Pennsylvania State University, \\ State College, PA, 16801}

\author[0000-0002-1989-3596]{Stuart D. Bale}
\affil{Physics Department, University of California, Berkeley, CA 94720-7300, USA}
\affil{Space Sciences Laboratory, University of California, Berkeley, CA 94720-7450, USA}

\author[0000-0001-6038-1923]{Kristopher Klein}
\affiliation{Lunar and Planetary Laboratory, University of Arizona, Tucson,
Tucson, AZ 85721, USA}

\author[0000-0002-7077-930X]{Justin Kasper}
\affiliation{BWX Technologies, Inc., Washington DC 20002, USA}
\affiliation{Climate and Space Sciences and Engineering, University of Michigan, Ann Arbor, MI 48109, USA}

\author[0000-0002-4401-0943]{Thierry {Dudok de Wit}}
\affil{LPC2E, CNRS and University of Orl\'eans, Orl\'eans, France}

\author[0000-0003-0420-3633]{Keith Goetz}
\affiliation{School of Physics and Astronomy, University of Minnesota, Minneapolis, MN 55455, USA}

\author[0000-0002-6938-0166]{Peter R. Harvey}
\affil{Space Sciences Laboratory, University of California, Berkeley, CA 94720-7450, USA}

\author[0000-0003-3112-4201]{Robert J. MacDowall}
\affil{Solar System Exploration Division, NASA/Goddard Space Flight Center, Greenbelt, MD, 20771}

\author[0000-0003-1191-1558]{David M. Malaspina}
\affiliation{Astrophysical and Planetary Sciences Department, University of Colorado, Boulder, CO, USA}
\affiliation{Laboratory for Atmospheric and Space Physics, University of Colorado, Boulder, CO, USA}

\author[0000-0002-1573-7457]{Marc Pulupa}
\affil{Space Sciences Laboratory, University of California, Berkeley, CA 94720-7450, USA}

\author[0000-0001-5030-6030]{Davin Larson}
\affiliation{Space Sciences Laboratory, University of California, Berkeley, CA 94720-7450, USA}

\author[0000-0002-0396-0547]{Roberto Livi}
\affiliation{Space Sciences Laboratory, University of California, Berkeley, CA 94720-7450, USA}

\author[0000-0002-3520-4041]{Anthony Case}
\affiliation{Smithsonian Astrophysical Observatory, \\
Cambridge, MA 02138, US}

\author[0000-0002-7728-0085]{Michael Stevens}
\affiliation{Smithsonian Astrophysical Observatory, \\
Cambridge, MA 02138, US}




\begin{abstract}
Parker Solar Probe (PSP) has shown that the solar wind in the inner heliosphere is characterized by the quasi omni-presence of magnetic switchbacks (``switchback'' hereinafter), local backward-bends of magnetic field lines. Switchbacks also tend to come in patches, with a large-scale modulation that appears to have a spatial scale size comparable to supergranulation on the Sun. Here we inspect data from the first ten encounters of PSP focusing on different time intervals when clear switchback patches were observed by PSP. We show that the switchbacks modulation, on a timescale of several hours, seems to be independent of whether PSP is near perihelion, when it rapidly traverses large swaths of longitude remaining at the same heliocentric distance, or near the radial-scan part of its orbit, when PSP hovers over the same longitude on the Sun while rapidly moving radially inwards or outwards. This implies that switchback patches must also have an intrinsically temporal modulation most probably originating at the Sun. Between two consecutive patches, the magnetic field is usually very quiescent with weak fluctuations. We compare various parameters between the quiescent intervals and the switchback intervals. The results show that the quiescent intervals are typically less Alfv\'enic than switchback intervals, and the magnetic power spectrum is usually shallower in quiescent intervals. We propose that the temporal modulation of switchback patches may be related to the ``breathing'' of emerging flux that appears in images as the formation of ``bubbles'' below prominences in the Hinode/SOT observations.

\end{abstract}



\section{Introduction} \label{sec:intro}
\begin{figure*}[htb!]
    \centering
    \includegraphics[width=\textwidth]{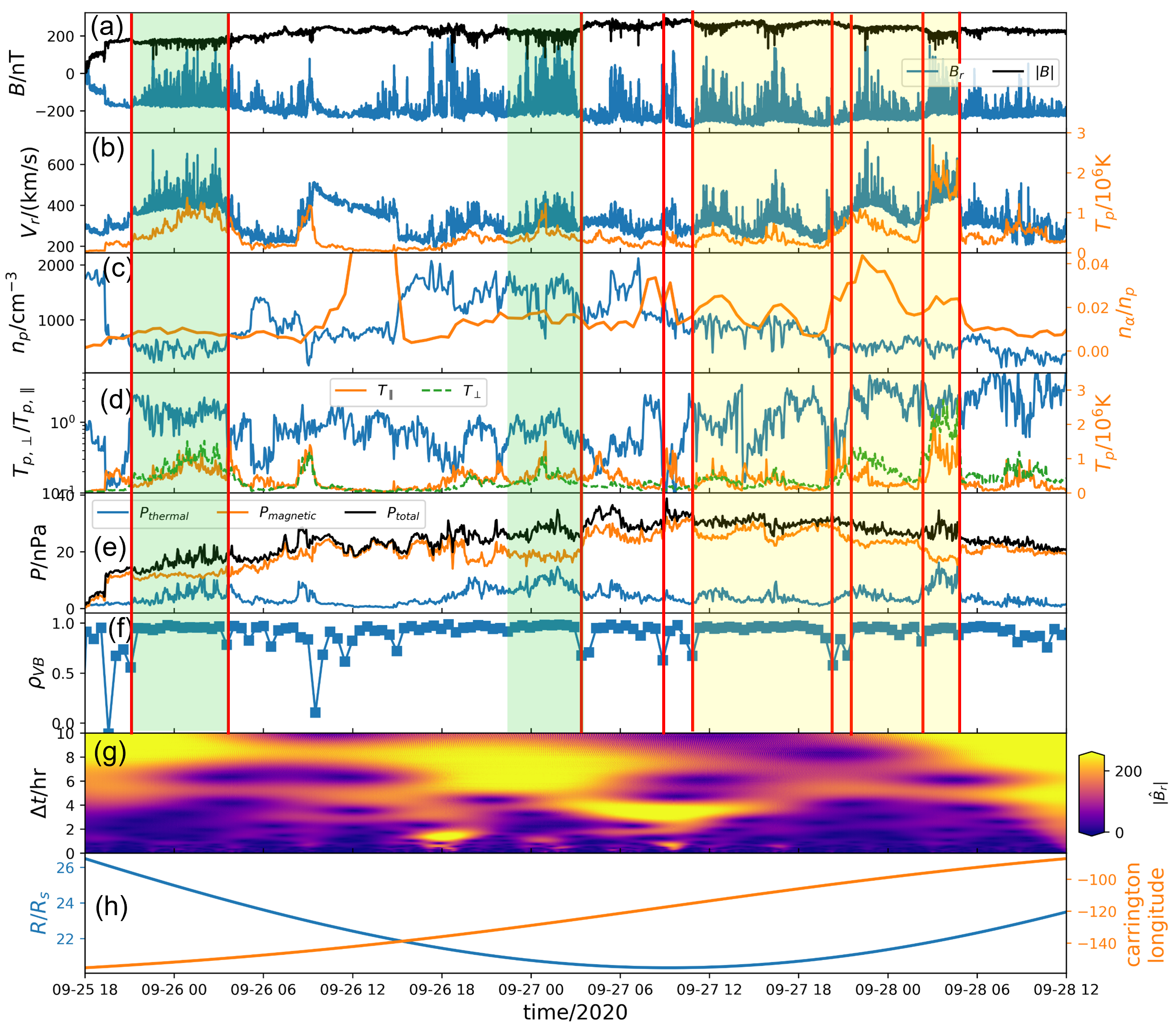}
    \caption{Time profiles of various quantities in the interval of interest (Sep. 25 18:00 - Sep. 28 12:00) during encounter 6. (a): Radial component (blue) and magnitude (black) of magnetic field. (b): Radial solar wind speed (blue) and proton temperature (orange). (c): Proton density (blue) and alpha particle abundance (orange) defined as the ratio between alpha and proton densities $n_\alpha/n_{p}$. (d): Proton temperature anisotropy $T_{p,\perp}/T_{p,\parallel}$ (blue), $T_{p,\parallel}$ (orange), and $T_{p,\perp}$ (green dashed). (e): Thermal pressure (blue), magnetic pressure (orange), and total pressure (black). (f): Alfv\'enicity $\rho_{VB}$ defined as the correlation between velocity and magnetic field fluctuations, calculated with half-hour time windows. (g): Wavelet transform of radial magnetic field. (h): Position of PSP. Blue curve is radial distance to the Sun, and orange curve is the Carrington longitude. Vertical red lines mark the dips in $\rho_{VB}$. Green and yellow shades correspond to the switchback patches discussed in the text. All the plasma data in this plot is from SPAN.}
    \label{fig:blowup_E06}
\end{figure*}

One of the most important findings made by Parker Solar Probe (PSP) is the large number of switchbacks in the near-Sun space \citep[e.g.][]{kasper2019alfvenic,bale2019highly}. They are local polarity-reversals of the radial magnetic field and they are typically highly Alfv\'enic \citep{larosa2021switchbacks}, with nearly constant magnetic field strength and highly correlated velocity and magnetic field fluctuations, corresponding to Alfv\'en waves propagating outwards from the Sun. 

\begin{figure}
    \centering
    \includegraphics[width=\hsize]{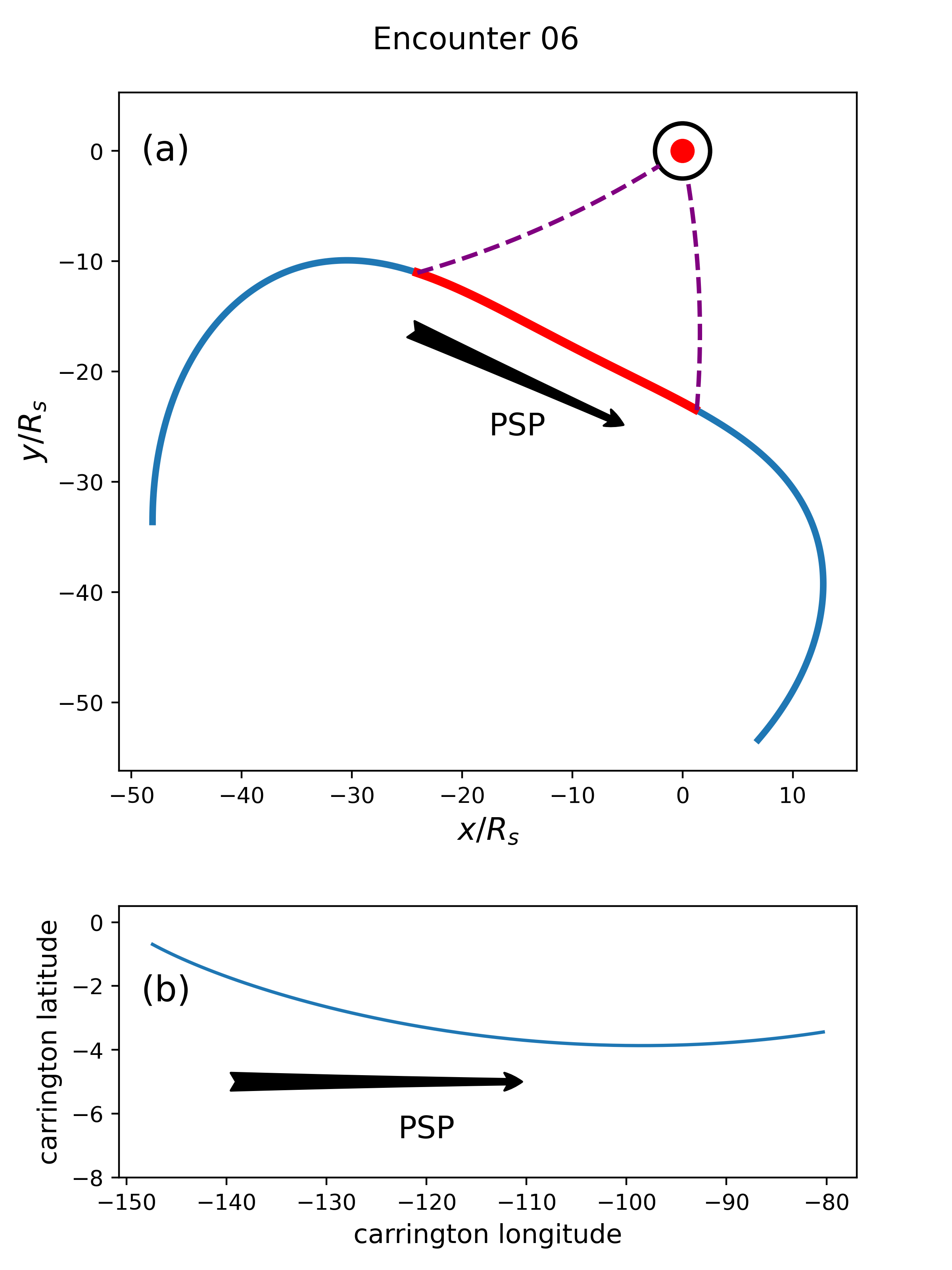}
    \caption{(a): Trajectory of PSP (blue curve) in encounter 6 projected on the solar equatorial plane. The reference frame is corotating with the Sun. The red circle represents the Sun, and the black circle represents the source surface ($r_{ss}=2.5R_s$). The red part of the trajectory is the time interval of the data shown in Figure \ref{fig:blowup_E06}. The two dashed curves show the ballistic projections of PSP to the source surface, using the average radial solar wind speed ($320$ km/s) in the interval marked by red. (b): Motion of the magnetic footpoint of PSP on the source surface derived by ballstic projection (the dashed curves in panel (a)) during the interval corresponding to the red part of the trajectory in panel (a). Horizontal and vertical axes are Carrington longitude and Carrington latitude respectively. In both panels, the black arrows indicate the direction of the PSP motion.}
    \label{fig:traj_E06}
\end{figure}

Many observational studies suggest that switchbacks play an important role in the evolution of the solar wind and the solar wind turbulence: turbulence properties are different inside and outside the switchbacks \citep{de2020switchbacks,bourouaine2020turbulence,martinovic2021multiscale}; \citet{hernandez2021impact} show that the switchbacks may inject additional energy in the turbulence cascade. At switchback boundaries, evidence of magnetic reconnection \citep{froment2021direct} and small scale wave activity \citep{mozer2020switchbacks} have been observed that may be dissipating the energy of the switchbacks. MHD simulations of a single switchback show that the switchback is eventually destroyed by parametric decay \citep{tenerani2020magnetic}, thus releasing energies to the background plasma. Indeed, there are studies showing that the proton temperature increases inside the switchbacks \citep[e.g.][]{mozer2020switchbacks,farrell2020magnetic}, though \citep{woolley2020proton} shows that the core proton parallel temperature does not show much difference inside and outside switchbacks.

Though many works analyze switchback characteristics \citep[e.g.][]{farrell2020magnetic,laker2021statistical,meng2022analysis}, the origin of switchbacks is still under debate, and several possible mechanisms have been identified. First, switchbacks may be a result of the evolution of the MHD turbulence in the expanding solar wind. MHD theory predicts that the amplitude of the Alfv\'enic magnetic filed fluctuations inside a radially expanding solar wind decays with radial distance as $\delta B \sim R^{-3/2}$ \citep[e.g.][]{hollweg1974transverse}. This means that the fluctuations decay slower than the radial component of the background magnetic field $B_r$ as $B_r \sim R^{-2}$. Therefore, the initially small fluctuation will eventually lead to the formation of a polarity reversal when its amplitude becomes larger than the background radial magnetic field. This in-situ generation mechanism has been confirmed by recent MHD simulations \citep{squire2020situ,shoda2021turbulent}. In addition,  interchange reconnection between closed and open fields in the solar corona \citep{zank2020origin,drake2021switchbacks} as well as velocity shears \citep{landi2006heliospheric,schwadron2021switchbacks}, and coronal jets \citep{sterling2020coronal} have been proposed as generating mechanisms. Though all of the above processes may be occurring, which one dominates in generating the switchbacks observed by PSP is still not clear. \citet{mozer2021origin} study the radial evolution of switchbacks and show that the hourly averaged count rate of switchbacks is quite constant at different radial distances to the Sun but increases with the solar wind speed. On the contrary, a statistical study conducted by \citet{tenerani2021evolution} shows that the count rates of switchbacks defined by the number of switchbacks per unit radial length the solar wind travels instead of unit time may either increase or decrease, depending on the scale, as the solar wind propagates outward, implying that switchbacks may be generated both locally in the solar wind and in the solar corona.

\citet{horbury2020sharp}, using encounter 1 data, show that the appearance and amplitude of the switchbacks are often modulated by a long time period of several hours. Recent studies by \citet{bale2021solar}, by mapping PSP to the solar corona, find that these switchback ``patches'' may correspond to spatial structures, the so-called magnetic ``funnels'', on the scale of supergranulation. Similar results are obtained by \citet{fargette2021characteristic}, who apply wavelet transform to the signal of radial magnetic field as a function of the Carrington longitude and find that the signal is strong around the typical angular sizes of supergranulation. However, the properties of the plasma inside the switchback patches seem to be complicated and non-universal. For example, for the patches observed in encounter 6 analyzed by \citet{bale2021solar}, the alpha particle abundance is higher in the patches than the quiescent wind, while \citet{woolley2021plasma} shows that the alpha particle abundance in an encounter 5 stream when SB patches were observed is lower compared with the surrounding quiescent wind. In addition, \citet{woodham2021enhanced} shows that inside the switchback patches observed in encounter 2, the parallel proton temperature is enhanced while the perpendicular proton temperature remains nearly constant, but \citet{bale2021solar} shows that the perpendicular proton core temperature is significantly enhanced in the switchback patches.

The PSP orbit crosses corotation at the beginning and end of each solar encounter. While at perihelion, PSP traverses large areas of the corona in longitude, at corotation crossing, the so-called fast radial scans, it moves radially while hovering over the same longitude on the Sun and extremely minor excursions in latitude. The persistence of switchbacks during fast radial scans would therefore provide evidence for a temporal component to their structure.

Solar magnetic activity expresses itself with spatial and temporal variability over multiple scales.  The magnetic field has been shown to 
concentrate on various spatial scales, from sunspots to ephemeral regions to supergranular structure to the  intranetwork field \citep{simon1964velocity, livingston1975new, harvey1973ephemeral, Gosic14}. The magnetic canopy formed in the chromosphere above \citep{reardon2011evidence} links the magnetically dominated corona to the photospheric field, with dynamic plasma behaviour seen in chromospheric fibrils and coronal cells. Chromospheric fibrils and coronal spicules are manifestations of the same structures, when observed on the disk and limb \citep{howard1964photospheric}. \citet{foukal1971morphological} show that in the chromosphere a filament channel may be recognized by the presence of fibrils aligned along the polarity reversal boundary. \citet{panasenco2010spicules} concludes that magnetic pattern of spicules/fibrils in the filament environment helps construct observation driven filament models and resolve an old puzzle of the bright rim observed near the feet and under solar filaments. Close relations between filament formation and supergranular cell dynamics were shown by \citet{su2012solar}.  Coronal cells \citep{sheeley2012coronal} originate from the network field concentrations, and have a supergranular scale and show the same pattern as chromospheric fibrils when inside filament channels. \citet{panasenco2014apparent} show that filament feet are anchored at the intersections of 4-5 supergranular cells.  \citet{berger2012quiescent} and \citet{berger2017quiescent} study dark bubbles appearance inside cold prominence observed by Hinode SOT with focus on the Rayleigh-Taylor instability (RTI) at their boundaries. However, the supergranular scale of these bubbles and their life time, together with their link to solar filament (prominences) allow us to compare these observable photospheric coronal properties with the in situ solar wind temporal and spatial parameters. Intersections of multiple supergranular cells naturally have stronger magnetic field concentration (a node), have a scale of supergranules and are locations above which chromospheric fibrils and coronal cells are centered. As shown in \citet{bale2021solar}, the magnetic network formed by these nodes plays a role in the spatial periodicities observed in SBs. 
Here we show that photospheric temporal properties also shape the observed SB patches seen by PSP.  

 We examine PSP data from first ten encounters to identify switchback patches. Specifically, we compare a selected time interval in encounter 6 with another two time intervals in encounters 1 and 10. We show that the switchback patches are not only associated with supergranulation-scale spatial structures but may also be related to some transient phenomena in the solar corona that lasts for several hours. The manuscript is organized as follows: In section \ref{sec:results}, we briefly introduce the data used in this study and show the PSP observations in selected time intervals. In section \ref{sec:discussion}, we discuss the observations and propose possible origins of the switchback patches. In section \ref{sec:conclusion}, we conclude this study.

\section{Results}\label{sec:results}
We use the level-2 fluxgate magnetometer data of the magnetic field vector from the FIELDS experiment of PSP with a time cadence of 3.4 milliseconds \citep{fox2016solar,bale2016fields}. For plasma measurements, we use the level-3 Solar Probe ANalyzer for Ions (SPAN-I) data with a time cadence of 7 sec for protons  and 14 sec for alphas, except for encounter 1 when high-quality SPAN data is unavailable and thus the level-3 Solar Probe Cup (SPC) data with highest time cadence of 0.44 sec is used \citep{fox2016solar,kasper2016solar}. We inspect data from the first ten encounters. For convenience, we will use ``E'' as abbreviation for ``encounter'', so that ``E01'' stands for ``encounter 1''.

\begin{figure}[htb!]
    \centering
    \includegraphics[width=\hsize]{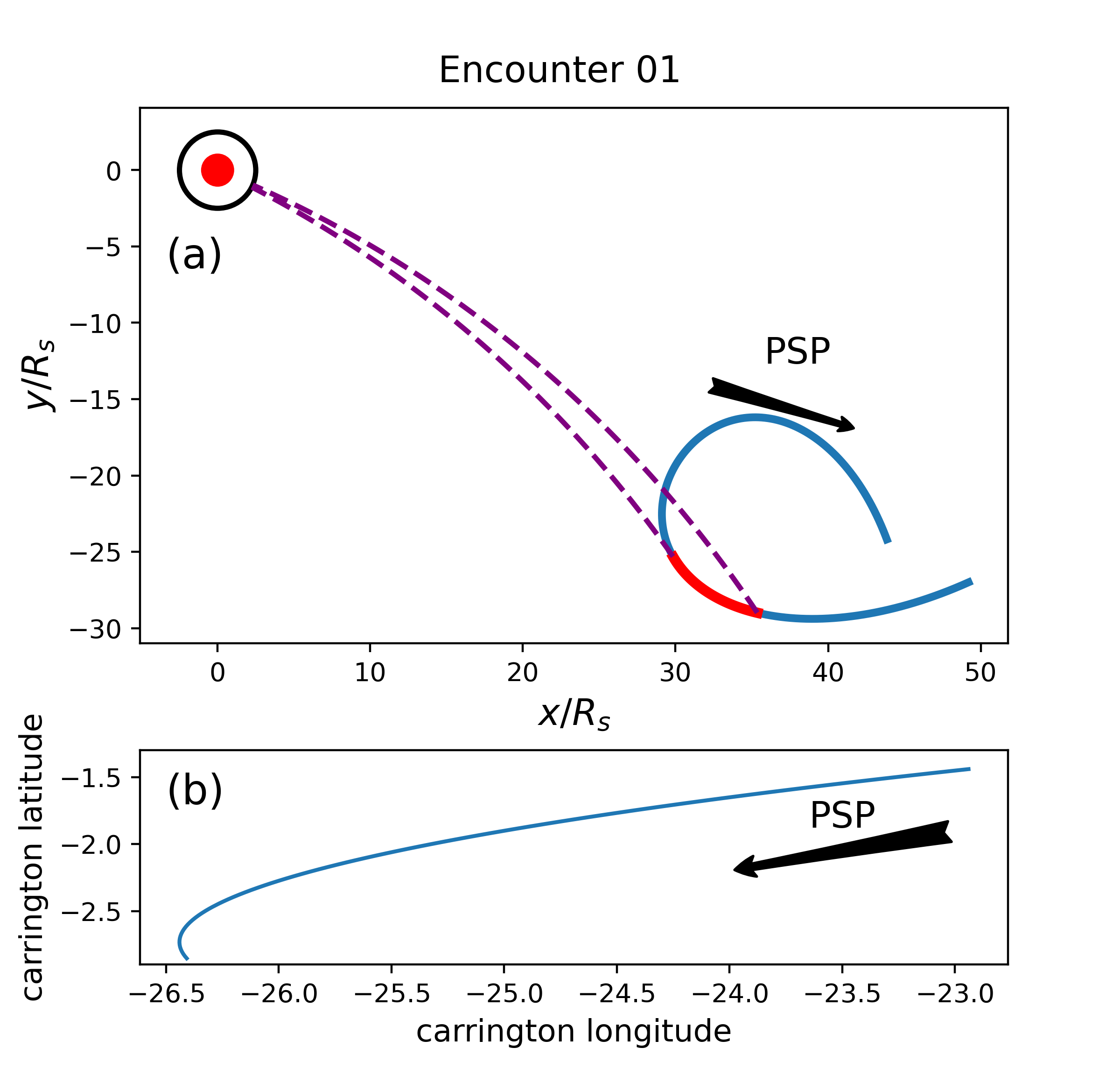}
    \caption{Similar to Figure \ref{fig:traj_E06} but for encounter 1. The average radial wind speed for the red interval is 280 km/s.}
    \label{fig:traj_E01}
\end{figure}

\subsection{Encounter 6 - spatial structures}\label{sec:encounter06}
We first analyze a time interval during E06, that is Sep. 25 18:00 to Sep. 28 12:00, 2020. This time interval was already analyzed by \citet{bale2021solar} but here we show additional diagnostics and we will compare this time interval with another two in E01 and E10, which will be discussed in section \ref{sec:encounter01}. In Figure \ref{fig:blowup_E06}, we plot time profiles of various quantities. Panel (a) shows the radial component (blue) and magnitude (black) of the magnetic field; Panel (b) shows the radial solar wind speed (blue) and proton temperature (orange); Panel (c) shows the proton density (blue) and alpha particle abundance (orange) defined as the ratio between alpha and proton densities $n_\alpha/n_{p}$; Panel (d) shows the proton temperature anisotropy $T_{p,\perp}/T_{p,\parallel}$ (blue), $T_{p,\parallel}$ (orange), and $T_{p,\perp} $ (green dashed); Panel (e) shows the thermal pressure (blue), magnetic pressure (orange), and total pressure (black); Panel (f) shows the Alfv\'enicity $\rho_{VB}$ defined as the correlation between velocity and magnetic field fluctuations, calculated with half-hour time windows: $\rho_{VB} = \left(\bm{u} \cdot \bm{b} \right) / \left( \left| \bm{u} \right| \left| \bm{b} \right|\right)$ where $\bm{u}$ and $\bm{b}$ are the fluctuations of velocity and magnetic field vectors; Panel (g) shows the wavelet transform of radial magnetic field so that we can read the dominant periods in $B_r$ signal; Panel (h) shows the position of PSP where blue curve is radial distance to the Sun, and orange curve is the Carrington longitude. All the plasma data in this plot is from SPAN. We acknowledge that SPAN-I only measures partial plasma moments and the uncertainties due to the obstructed field-of-view (FOV) are still under investigation. During the interval shown in this figure, SPAN-I has sufficient FOV to make an assessment on this timescale.

\begin{figure*}[htb!]
    \centering
    \includegraphics[width=\textwidth]{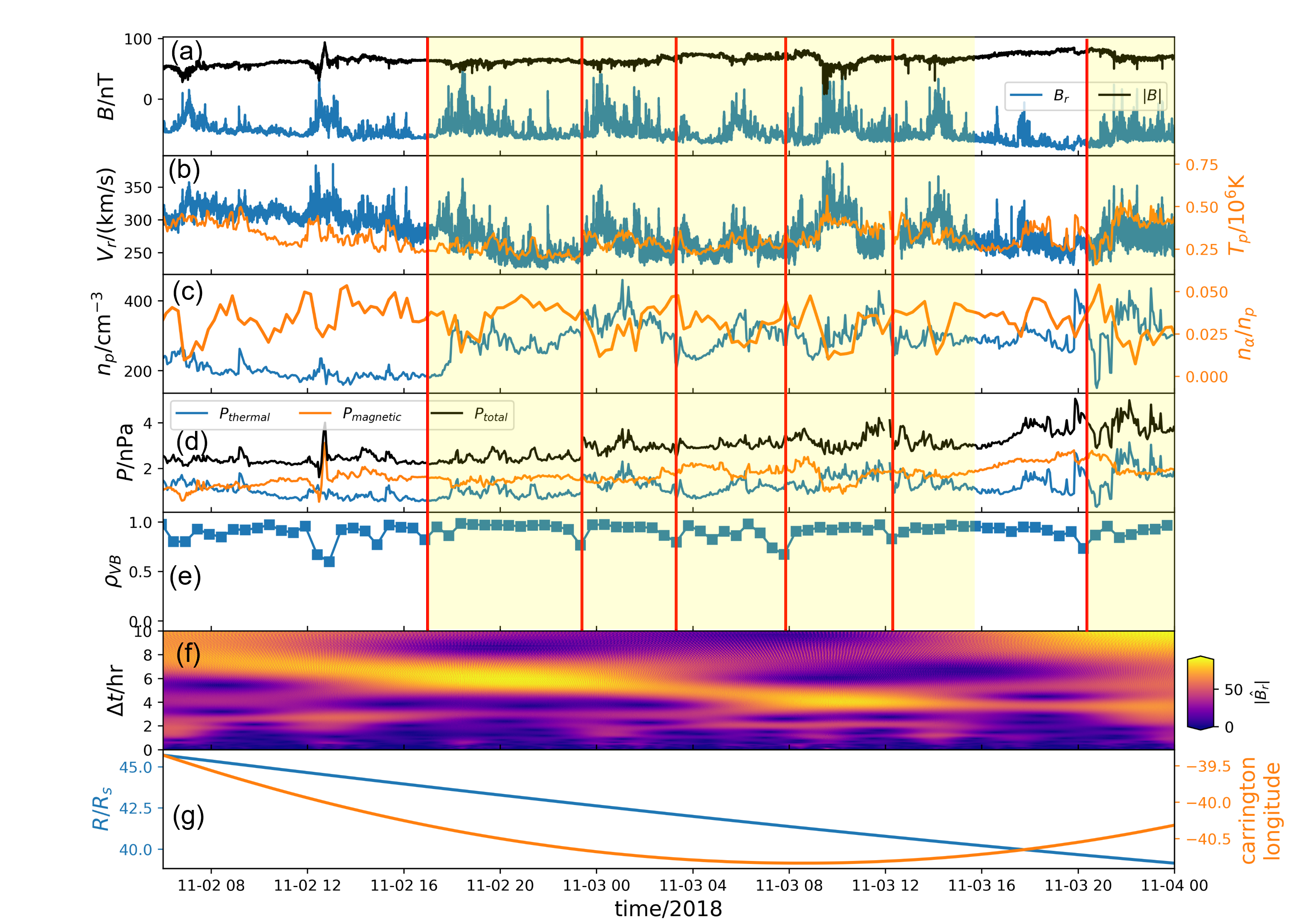}
    \caption{Similar to Figure \ref{fig:blowup_E06}, but for Nov. 02 06:00 - Nov. 04 00:00, 2018, during encounter 1. Because there is no high-quality SPAN data, SPC data was used for all the plasma parameters in this figure and hence the plot of temperature anisotropy is missing.}
    \label{fig:blowup_E01}
\end{figure*}

We can see four repeating switchback patches between 09/27 11:00 and 09/28 05:00, marked by the yellow shade, and they have the following characteristics: (1) The occurrence and amplitude of switchbacks are modulated with a time period of several hours, about four to six hours as can be read from panel (g). (2) The base of $B_r$ is modulated so that $|B_r|$ decreases at the center of each patch. (3) The magnetic field strength decreases in the patches. (4) The (large-scale) radial solar wind speed and the proton temperature increase in the patches. (5) The alpha particle abundance is enhanced in the patches. (6) The total pressure is roughly conserved, implying a pressure-balance structure. The magnetic pressure drops and thermal pressure increases in the switchback patches. (7) $T_{p,\perp}/T_{p,\parallel}$ is in general larger, with values greater than unity, in the patches. (8) The patches are asymmetric with a sharp leading edge and smooth trailing edge. From these observations, \citet{bale2021solar} inferred that the switchbacks originate from the magnetic ``funnels'' expanding out from the network between supergranules, and the switchback patches correspond to microstreams ejected from these regions as the enhanced wind speed and alpha particle abundance indicate a fast type of solar wind stream. The drop in magnetic field strength is due to the super-radial expansion of the magnetic funnels. The asymmetry of the patches is a longitudinal effect because of the differential rotation between the photosphere and solar corona or some inherent asymmetry of the funnel structure. The enhanced temperature anisotropy implies an overall stronger perpendicular heating possibly due to a stronger turbulence dissipation inside the switchback patches. However, we note that panel (d) shows $T_{p,\perp}$ is quite well-correlated with the switchback patches while $T_{p,\parallel}$ is more irregular. This irregular $T_{p,\parallel}$ leads to sharp increases and decreases of temperature anisotropy (e.g. the sharp jump around 09/27 22:30). Besides, for the patch around 09/27 12:00, the $T_{p,\perp}/T_{p,\parallel}$ ratio drops at the center of the patch due to a larger increase in $T_{p,\parallel}$ than $T_{p,\perp}$. Thus, the behavior of temperature anisotropy is quite complicated and its correlation with the switchback patches is not as well defined as parameters like the temperature. From panel (f), we can see that the Alfv\'enicity is in general very high, especially inside the switchback patches, with values close to 1. But there are also some significant drops, which are marked by the vertical red lines. One can see that the drops in $\rho_{VB}$ are mostly found immediately before or after the switchback patches, i.e. in the quiescent intervals.

Besides the four repeating patches, several other patches can also be identified. They have some similar characteristics with those discussed in the prior paragraph, including the similar time scales, increased $V_r$ and $T_p$, depressed $|\mathbf{B}|$, enhanced $T_{p,\perp}/T_{p,\parallel}$, and low-Alfv\'enicity periods bounding them. However, there are also some discrepancies. Take the the two patches marked by the green shades as examples. First, we can see that the alpha particle abundance only shows very slight enhancement at their centers. Second, the asymmetry of the two patches is opposite to those marked by the yellow shade, i.e. they have a smooth leading edge and a sharp trailing edge, while throughout the whole time interval the Carrington longitude of PSP is monotonically increasing, implying that the asymmetry may not be explained by the differential rotation of the photosphere and solar corona because the differential rotation predicts that sharper edge of the switchback patch should always be at the smaller longitude side \citep{bale2021solar}.

In panel (a) of Figure \ref{fig:traj_E06}, we show the trajectory of PSP in E06 projected on the solar equatorial plane (blue curve). In this panel, the red circle is the Sun and the black circle is the location of source surface at $r_{ss}=2.5R_s$ where $R_s$ is the solar radius. The red part of the trajectory is the time interval corresponding to Figure \ref{fig:blowup_E06}, which is around the perihelion of PSP's orbit. The two dashed curves show the ballistic projections of PSP to the source surface using the averaged solar wind speed (320 km/s) in the interval marked by red. In panel (b), we show the motion of the magnetic footpoint of PSP on the source surface derived by ballistic projection during the time interval marked by red in panel (a). In this panel, the horizontal axis is Carrington longitude and the vertical axis is Carrington latitude. We can see that, during this time interval of 66 hours, PSP traveled from -150 to -80 degrees in longitude. Hence, assuming these switchback patches correspond to some spatial structures in the solar corona, we can roughly estimate that each individual patch, which usually lasts for 4-6 hours, is about 4-6 degrees in longitude, similar to the estimate by \citet{bale2021solar}. This angular size is close to the typical size of supergranules.

\subsection{Encounters 1 \& 10 - temporal structures}\label{sec:encounter01}
Because the perihelion of PSP's orbit is very low, the angular speed of PSP exceeds the angular speed of solar rotation near the perihelion. Therefore, in the corotating reference frame of the Sun, PSP travels to the west as it approaches the perihelion, then to the east around the perihelion, and finally to the west again as it moves away from the perihelion (see Figure \ref{fig:traj_E06}). Consequently, there are two corotating intervals in each orbit, when PSP moves in an almost purely radial direction in the corotating reference frame of the Sun. This unique feature has made it possible to observe the solar wind stream from a particular solar source region twice during one perihelion at two different radial locations \citep{shi2021alfvenic}. In addition, around the corotating intervals, PSP continuously measures the same stream for a long period.

In Figure \ref{fig:traj_E01}, we plot the trajectory of PSP in encounter 1 in a similar way with Figure \ref{fig:traj_E06}. We note that the perihelion of encounter 1 is around 35 $R_s$, higher than encounter 6 whose perihelion is around 20 $R_s$. We mark the inbound corotating interval with red in panel (a) and applied a ballistic projection from PSP to the source surface as shown by the dashed curves. In panel (b) the motion of the magnetic foot point on the source surface is plotted. We can see that, because PSP is almost moving radially in the corotating reference frame, its foot point moves only slightly in Carrington longitude, about 3.5 degrees within 42 hours.

In Figure \ref{fig:blowup_E01}, we plot time profiles of various quantities during the inbound corotating interval (Nov. 02 06:00 - Nov 04 00:00), i.e., the red part of the trajectory shown in Figure \ref{fig:traj_E01}. We note that there is no exact definition of the corotating interval, so the choice of the interval length is arbitrary. Because of the lack of high-quality SPAN data in E01, we use SPC data for all the plasma parameters and there is no temperature anisotropy data. We can see that, during this time interval, several switchback patches were observed by PSP, as marked by the yellow shades. Depression in $|\mathbf{B}|$, increase in $V_r$, and decrease in $\rho_{VB}$ before and after each patch are clearly seen, similar to the E06 observation shown in Section \ref{sec:encounter06}. The proton temperature is positively correlated with the switchback patches during some time intervals, e.g. 11/03 04:00 - 11/03 12:00, but drops inside other patches, e.g. around 11/03 00:00 and 11/03 14:00. But in E06 the correlation is well positive (Figure \ref{fig:blowup_E06}). In addition, from panel (c), one can see that the alpha particle abundance drops inside the patches, with a very strong negative correlation with the switchback amplitude shown in panel (a). This result is in contrast to the E06 observation. Panel (f) shows that the duration of each switchback patch is about 4-6 hours, very close to those observed in E06. However, as discussed before, PSP was almost corotating with the Sun, and its magnetic foot point on the source surface only moved by about 3.5 degrees throughout this time interval. We can roughly estimate the angular size of each switchback patch is about 0.3-0.5 degree, which is one order of magnitude smaller than the angular size estimated in the previous section and thus it is much smaller than the typical size of supergranules. 

\begin{figure}[htb!]
    \centering
    \includegraphics[width=\hsize]{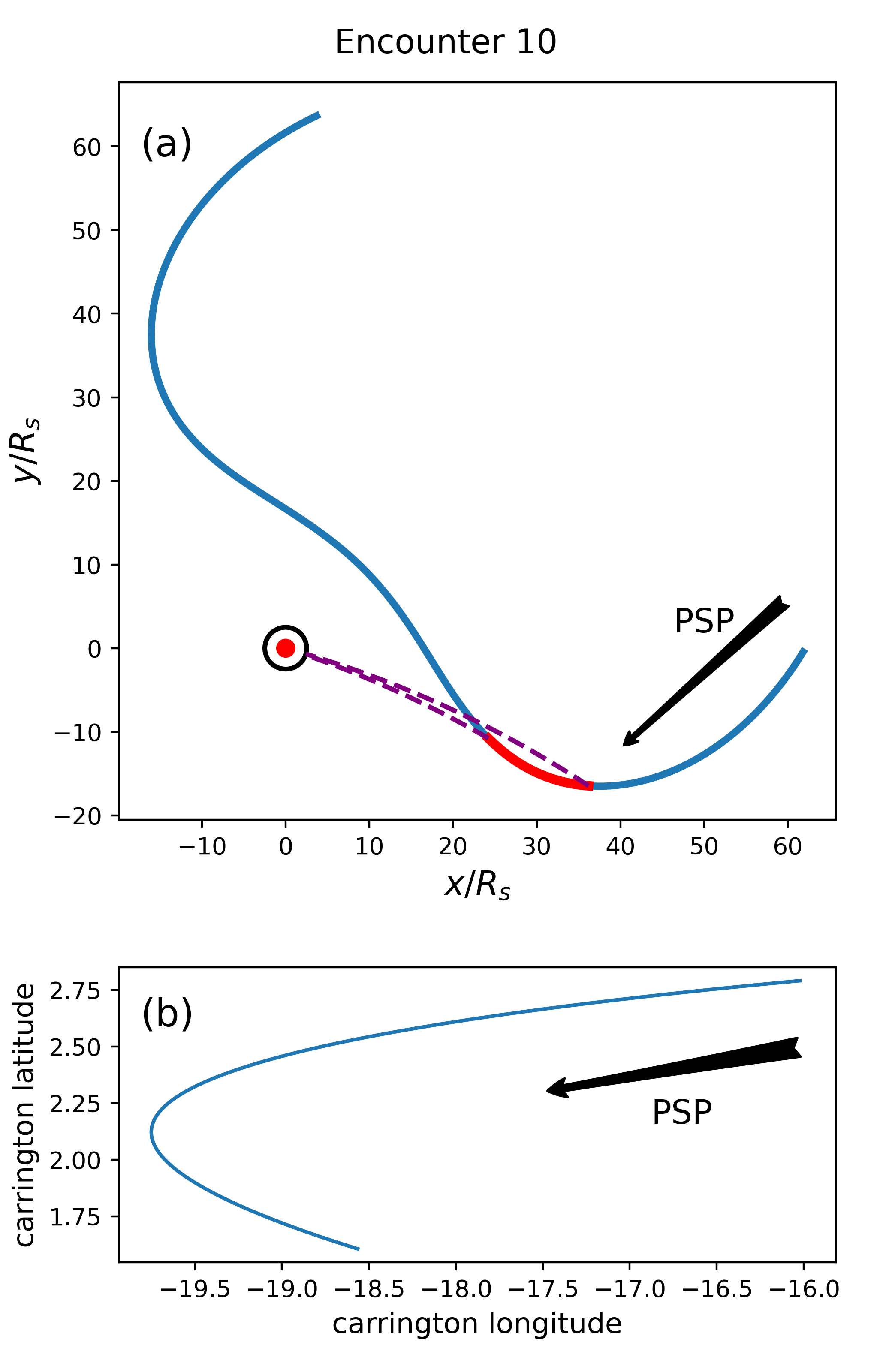}
    \caption{Similar to Figure \ref{fig:traj_E06} but for encounter 10. The average radial wind speed for the red interval is 470 km/s.}
    \label{fig:traj_E10}
\end{figure}

\begin{figure*}[htb!]
    \centering
    \includegraphics[width=\textwidth]{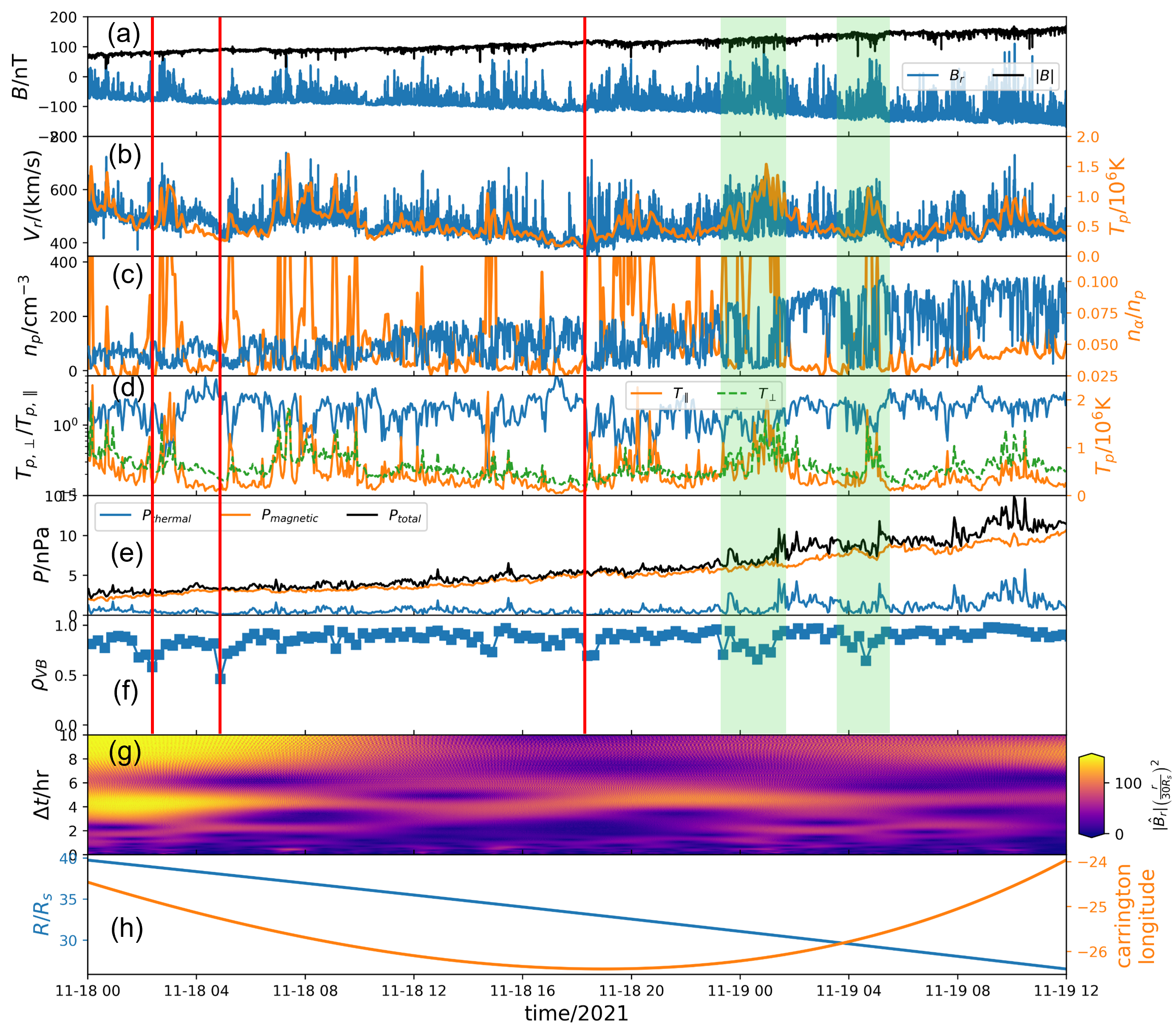}
    \caption{Similar to Figure \ref{fig:blowup_E06}, but for Nov. 18 00:00 - Nov. 19 12:00, 2021, during encounter 10.}
    \label{fig:blowup_E10}
\end{figure*}

In Figure \ref{fig:traj_E10} and Figure \ref{fig:blowup_E10}, we show the trajectory of PSP and the time profiles of different quantities around the corotating interval in E10. Panel (a) of Figure \ref{fig:blowup_E10} clearly shows that there are repeated switchback patches throughout this time interval. From panel (g), we can observe a strong signal around 4-5 hours, similar to the E01 observation (Figure \ref{fig:blowup_E01}), implying a similar origin of the patches for the two intervals. In addition, drops in Alfv\'enicity are observed in the quiet streams ahead of some switchback patches, as marked by the vertical red lines. We note that the Alfv\'enicity in this interval is in general slightly lower than the E06 and E10 intervals, though it is still quite high ($\rho_{VB} \approx 0.7-0.9$). The reason is not clear but is possibly related to the influence of the solar wind sources. In contrast to E06 observation (Figure \ref{fig:blowup_E06}), $T_{p,\perp}/T_{p,\parallel}$ decreases inside most of the switchback patches during this interval. Additional analysis is necessary to fully understand the behavior of the proton temperature anisotropy in the switchback patches, but we note that the assessment of proton temperature depends on if one is analyzing the full velocity distribution function or the separate beam/core components \citep[e.g.][]{klein2021inferred}. Additional caveats related to SPAN-ion field-of-view obstruction must also be considered, such as the accuracy constrained by partial velocity distribution function measurements. Another interesting point in Figure \ref{fig:blowup_E10} is that, there are several low-Alfv\'enicity intervals inside the switchback patches, such as those shaded by green. From panel (c), we can see that these two intervals are accompanied with significant drops in proton density. Thus, the two switchback patches may have quite different origins compared with others. A detailed analysis of them is important but will be carried out in future studies.

To summarize, the observations in the selected corotating intervals of encounters 1 \& 10 indicate that the observed switchback patches are possibly not related with the spatial scale of supergranules, as PSP almost stayed at a constant Carrington longitude. Instead, since the duration of each individual patch is quite similar (4-6 hours) among the two encounters (and even E06), it is likely that they correspond to some temporal phenomenon happening in the solar corona, which will be discussed in more details in section \ref{sec:discussion}.  


\subsection{Switchback intervals and quiescent intervals}\label{sec:sb_and_quiet}

\begin{figure*}[htb!]
    \centering
    \includegraphics[width=\textwidth]{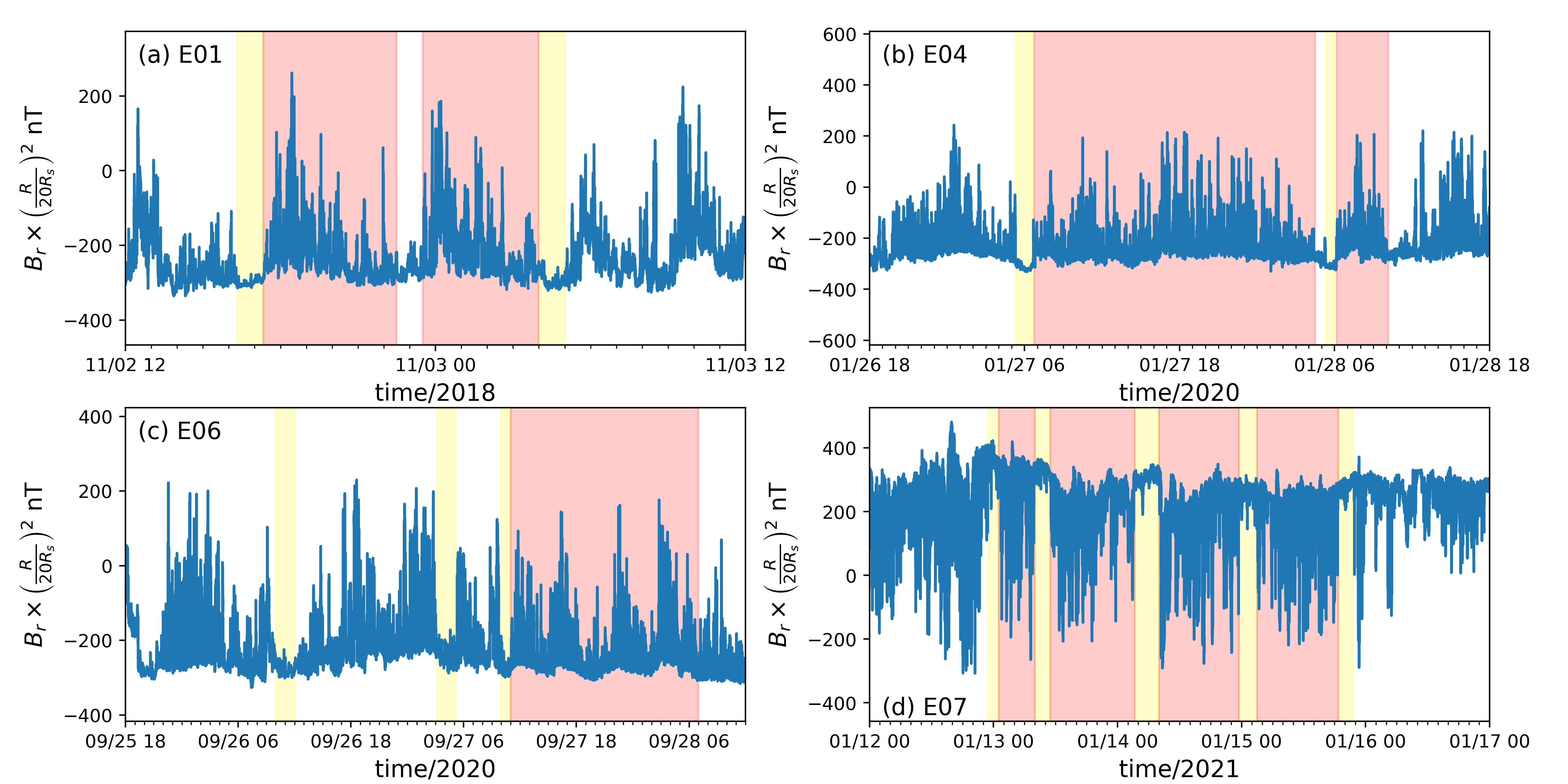}
    \caption{Time profiles of $B_r$ normalized by $(R/20R_s)^2$ for selected intervals in encounters 1, 4, 6, and 7. In each panel, the yellow shades mark the extremely quiescent patches and the red shades mark the switchback patches.}
    \label{fig:quiet_sb_Br_time}
\end{figure*}

\begin{figure*}[htb!]
    \centering
    \includegraphics[width=\textwidth]{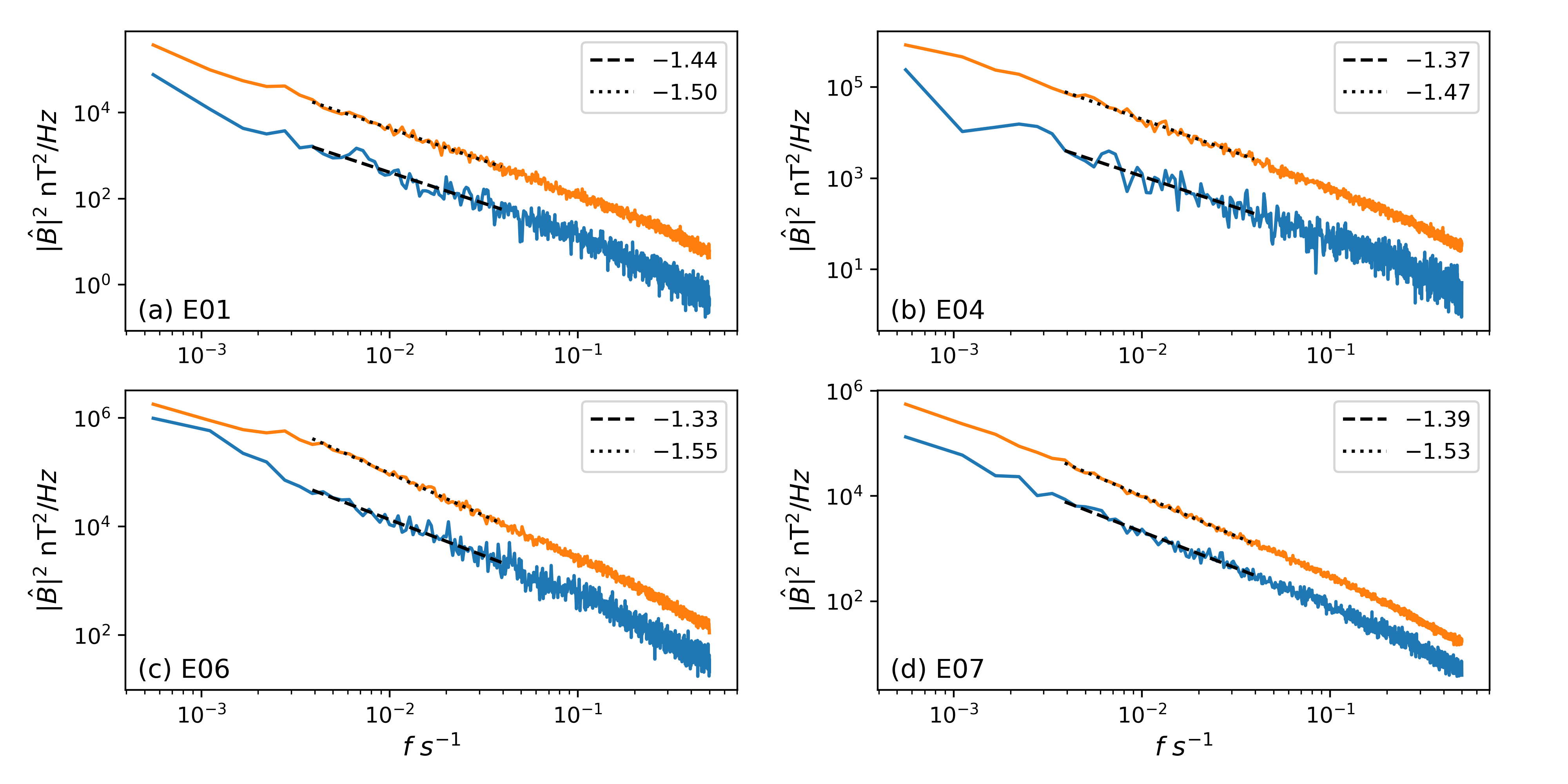}
    \caption{Power spectra of magnetic field for selected time intervals in encounters 1, 4, 6, and 7, as shown in Figure \ref{fig:quiet_sb_Br_time}. In each panel, the blue curve is the spectrum for quiescent intervals, and the orange curve is the spectrum for switchback intervals. The dashed and dotted lines are the linear fits of the two spectra and their slopes are written in the legends.}
    \label{fig:quiet_sb_spectrum}
\end{figure*}

As shown before, streams between the switchback patches are typically extremely quiescent, with very weak fluctuations. In Figure \ref{fig:quiet_sb_Br_time}, we plot $B_r$ multiplied by $(R/20R_s)^2$ for selected intervals in encounters 1, 4, 6, and 7. Here, the E01 (corotating) and E06 (perihelion) intervals are those shown in Figures \ref{fig:blowup_E01}\&\ref{fig:blowup_E06}, the E04 interval is after the inbound corotating interval and before the perihelion, and the E07 interval is around the inbound corotating interval. In each panel, the yellow shades mark the quiescent patches and the red shades mark the switchback patches. These quiescent patches are much shorter than the switchback patches and their duration is usually between half hour and three hours. In Figure \ref{fig:quiet_sb_spectrum}, we plot the power spectra of $B_r$ calculated for the quiescent (blue) and switchback (orange) intervals corresponding to the yellow and red shades in Figure \ref{fig:quiet_sb_Br_time}. For each encounter, we apply a Fourier transform to each individual quiescent (switchback) interval, then we re-sample each spectrum to an array of Fourier modes whose longest and shortest wave periods are 30 minutes and two seconds respectively, and finally we average the spectra if there are more than one quiescent (switchback) intervals. The power level is naturally lower in the quiescent intervals than the switchback intervals due to the small amplitude of the fluctuations. We linear-fit the spectra over the frequency range $f\in[4\times 10^{-3},4  \times 10^{-2}] s^{-1}$ which is well inside the MHD inertial range. The fitted spectra are plotted as the dotted and dashed lines in the figure and their slopes are written in the legends. In the switchback intervals the power spectra are in general steeper, with slopes around -1.5, while in the quiescent intervals the spectra are shallower, with slopes around -1.4. This result is qualitatively consistent with the results of previous studies \citep{de2020switchbacks,bourouaine2020turbulence}, but some discrepancies are worth noting here. The methods to calculate the power spectra are different among the studies. \citet{de2020switchbacks} adopt a Lomb-Scargle method to estimate the power spectral density from irregularly sampled data and obtain time scales exceeding the duration of individual quiescent intervals. \citet{bourouaine2020turbulence} use the conditioned correlation functions and thus can also estimate the power spectra from irregularly distributed data points. Compared with these two studies, the power spectra we get are calculated based on continuous time intervals. The quiescent intervals in \citep{de2020switchbacks,bourouaine2020turbulence} mainly refer to the intervals between individual switchbacks while in our work they refer to the prolonged time periods between switchback patches. The different methodologies may lead to the different spectral slopes acquired among the three studies: In \citep{bourouaine2020turbulence} the slopes of magnetic field spectra are -1.51 in quiescent intervals and -1.74 in switchback intervals, and in \citep{de2020switchbacks} the two slopes are -1.07 and -1.50 respectively. Interestingly, another study by \citet{martinovic2021multiscale} shows that the spectral slope in the inertial range is quite constant inside and outside a single switchback. But they only analyzed one switchback, so whether this result is statistically robust is a question. It seems that the power spectra are in general steeper in the switchback intervals, indicating that switchbacks play an important role in the turbulence energy cascade.


\begin{figure*}[htb!]
    \centering
    \includegraphics[width=\textwidth]{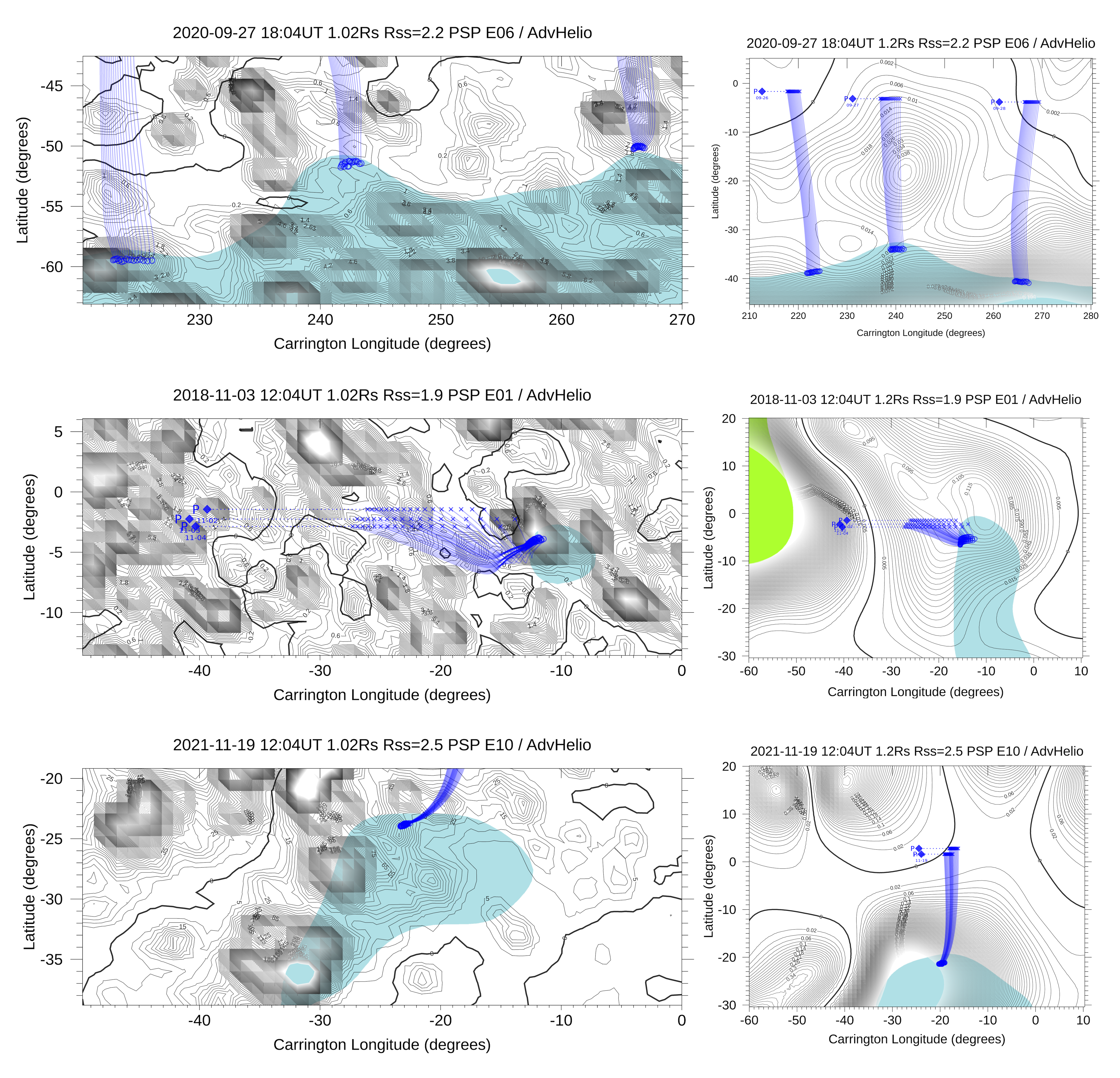}
    \caption{Magnetic field connectivity with the solar sources during PSP Encounter 06 (top row), Encounter 01 and 10 (middle and bottom rows). The thick black lines are the model neutral lines. Black contours indicate magnetic field pressure at 1.02 Rs ($\sim$ 14 Mm) (left column) and at 1.2 Rs (right column). The ballistic projection of the PSP trajectory (blue diamonds) on the source surface (blue crosses) and down to the solar wind source regions (blue circles) is calculated for different source surfaces $R_{ss}/R_s$=2.2, 1.9, and 2.5 from top to bottom (see \citet{panasenco2020exploring} for details) and measured in-situ solar wind speed $\pm$ 80 km/s. Open magnetic field regions are shown in blue (negative) and green (positive). 
}
    \label{fig:sources}
\end{figure*}

\begin{figure*}[htb!]
    \centering
    \includegraphics[width=\textwidth]{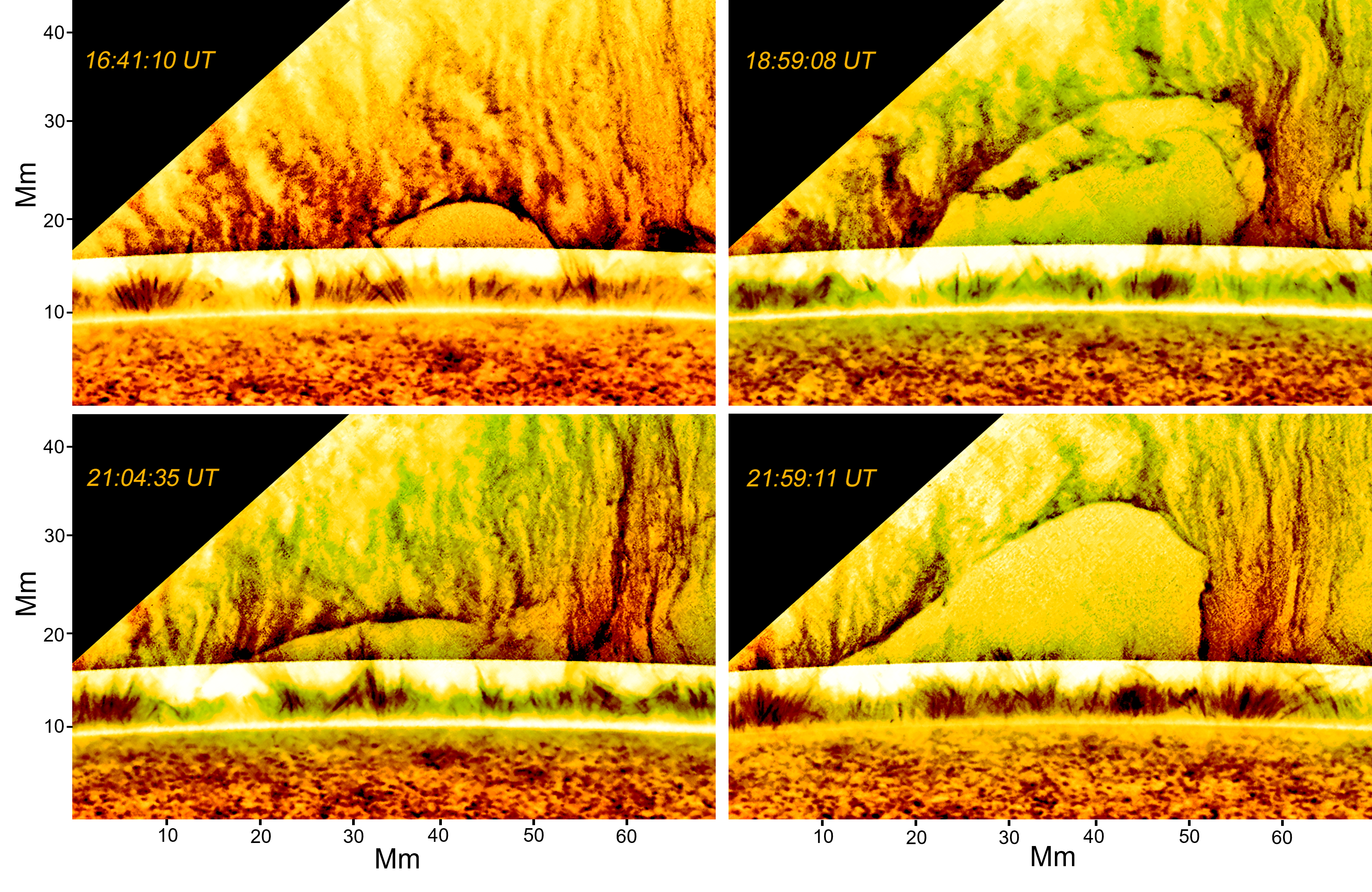}
    \caption{Prominence and spicules observed by the Hinode SOT in the CaH spectral line on August 16, 2007. The width of the bubbles is 40 - 45 Mm, the height reaches 30-35 Mm. The evolution of the one given bubble takes about 4 hours. The top panels show the same bubble as it growing and collapsing soon after 19:30 UT. However a new bubble takes its place, with a bit faster growing rate. The spatial scale of bubbles and the fact that prominence footpoints are anchored at the intersection of multiple supergranular cells allows us to link apparent bubble properties and location to supergranules. Supergranules life time in average about 24 hours, which include all stages of their evolution including the emergence, spatial grow and decay. Second bubble (bottom row) evolved faster and emerged already with the width of a well developed previous bubble, this fact allows us to suggest that during the life of one supergranule we can observe 4-5 cycles of bubble development.}
    \label{fig:SOT}
\end{figure*}

\begin{figure*}[htb!]
    \centering
    \includegraphics[width=\textwidth]{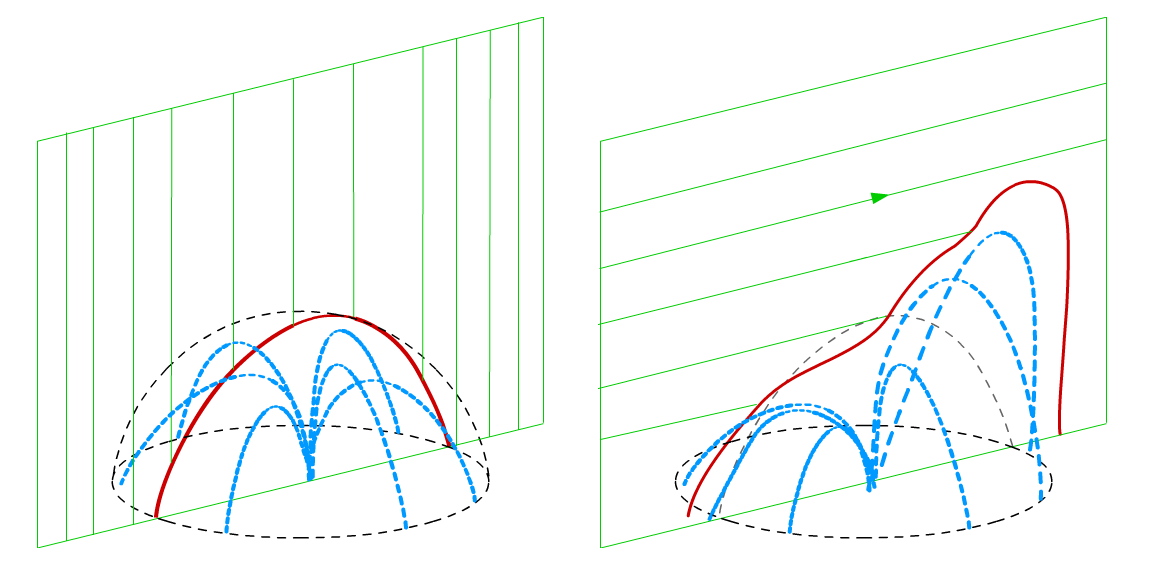}
    \caption{The differences in supergranular magnetic flux dome/bubble when emerged into vertical and horizontal magnetic field. The left panel show a symmetrical dome profile (red line); the right panel represents bubble flux emergence into the solar prominence plasma sheet with a strong horizontal component of the magnetic field, resulting in a very asymmetrical shape of the prominence bubbles as shown in the right column Figure \ref{fig:SOT}. }
    \label{fig:dome}
\end{figure*}

\section{Discussion}\label{sec:discussion}
\subsection{Source regions of switchback patches in the solar corona}
Previous studies \citep{bale2021solar,fargette2021characteristic}, by mapping the structure back to the solar corona, find that the angular size of these switchback patches is typically on the order of several degrees, coincident with the size of supergranules. In addition, the switchback patches observed in E06 show features of typical fast stream such as larger wind speed and larger alpha particle abundance \citep{bale2021solar}, implying that they may originate from the magnetic funnels on top of the supergranules. However, the observations in E01 \& E10 (Section \ref{sec:encounter01}) seem to provide a different picture, as obviously repeated switchback patches were observed near the corotating interval of PSP's orbit, when PSP as well as its magnetic footpoint on the source surface moved extremely slowly in Carrington longitude. 

In Figure \ref{fig:sources}, we show the magnetic footpoints of PSP at the base of solar corona for the three time intervals analyzed in Section \ref{sec:results}. From top to bottom rows are E06, E01, and E10 respectively. The left column shows the magnetic pressure maps calculated for the surface at 1.02$R_s$, and right column shows the magnetic pressure maps calculated for the surface at 1.2$R_s$. We note that the left column displays smaller regions than the right column. In each panel, the colored regions are the open magnetic field regions. The blue diamonds are the direct radial projections of PSP on different dates. The blue crosses are the magnetic footpoints of PSP on the source surface after a ballistic projection. The source surface height is written in the title of each plot. For each ballistic projection, we have 17 blue crosses that correspond to a series of varying solar wind speeds (a cross for speed measured in situ by PSP and $\pm$80 km/s with 16 more crosses for every 10 km/s increment). The blue circles are the footpoints at the base of solar corona that connect to the blue crosses through the potential field source surface (PFSS) model \citep{altschuler1969magnetic,schatten1969model}. From the top row, one can see that in the E06 interval, the source region of the solar wind measured by PSP was located at the boundary of the southern polar coronal hole, and it traveled very fast in longitude. In contrast, the middle and bottom panels show that during the E01 \& E10 intervals, the source regions almost stayed stationary. For the E01 interval, the source region was located at the boundary of an equatorial coronal hole, while for the E10 interval, the source region was located at the boundary of the equatorial extension of the southern polar coronal hole. Thus, the very tiny amount of the footpoint motion in E01 \& E10 excludes the possibility that these switchback patches correspond to spatial structures on the scales of supergranules.


The above observations imply the possibility that the switchback patches observed in the E01 and E10 corotating intervals are generated as a result of some temporal processes in the solar corona. That is, because of a certain mechanism, the magnetic field and plasma in the source region of the slow solar wind measured by PSP oscillate at periods of several hours. During the oscillations, fluctuations are injected to the solar wind streams, and either these fluctuations are already strong enough to reverse the radial magnetic field, or they eventually evolve into switchbacks as the solar wind expands radially \citep[e.g.][]{squire2020situ,shoda2021turbulent}. During time intervals when the oscillation at the solar wind source is weak, few fluctuations are injected and these fluctuations are less Alfv\'enic. Besides, because of the weak fluctuation level and thus weak nonlinear interactions, the power spectrum of magnetic field is overall shallower in the quiescent intervals as shown in Figure \ref{fig:quiet_sb_spectrum}. The uncertain correlation between the alpha particle abundance and the switchback patches (by comparing Figure \ref{fig:blowup_E06} and Figure \ref{fig:blowup_E01}) is mysterious. One possibility is the different base levels of $n_\alpha/n_p$. One can see that, toward the centers of the switchback patches, in E06 $n_\alpha/n_p$ increases from about 0.01 to 0.02-0.04 while in E01 $n_\alpha/n_p$ drops from 0.05 toward 0.025. The base level of the alpha particle abundance in E01 is quite high and close to the fast solar wind level. We note that, the sources of the solar wind in E01 and E06 are different (see Figure \ref{fig:sources}). In E01 the solar wind originates from an equatorial coronal hole \citep{panasenco2020exploring}, while in E06 the solar wind originates near the polar coronal hole boundaries \citep{bale2021solar}. Hence, it is possible that the mechanism that generates the switchback patches tends to produce a similar alpha particle abundance around 0.02-0.04 within different background plasma environments. However, we note that the alpha particle abundance in E10 is highly fluctuating (Figure \ref{fig:blowup_E10}), possibly due to the fluctuations of proton density, and no clear correlation between the alpha particle abundance and the switchback patches is observed.

\subsection{Magnetic nature of prominence bubbles}
Prominence bubbles \citep{berger2010quiescent,berger2011magneto,berger2017quiescent} are hot underdense plasma cavities seen to expand from the solar surface into prominences in HINODE/SOT images. Figure \ref{fig:SOT} shows four consecutive CaH-line pictures taken by the Hinode Solar Optical Telescope (SOT) on August 16, 2007. The top row shows the growth of one bubble and the bottom row shows the growth of a subsequent bubble at the same location. The horizontal size of the bubbles is around 35-40 Mm, i.e., similar to the typical supergranule size of 30-40 Mm. Bubbles grow before breaking down through apparent instabilities over a timescale of  3-6 hours. These bubbles may exist everywhere at the bottom of the solar corona, but are observable only when they form and grow under solar prominences thanks to contrast with the prominence plasma material. Bubble collapse is most probably associated with Rayleigh-Taylor instabilities developing once field lines are long enough to reduce the line-tying stabilizing effect \citep{PV09} at heights comparable to their diameter.

The Hinode/SOT images do not provide magnetic properties of the bubbles directly. However, the magnetic nature of bubbles observed inside thin layers of solar filament/prominence plasma sheets can be inferred from the way these dark domes interact with the surrounding magnetic field.  Figure \ref{fig:SOT} shows an example of dome evolution below and inside a quiescent prominence: once the upper dome boundary reaches a certain height, the dome shape becomes asymmetric, a feature that may be attributed to interaction with the strong horizontal component of the prominence axial magnetic field, in a way similar to how chromospheric fibrils and coronal cells - overlying the supergranular network -  manifest their magnetic nature \citep{panasenco2010spicules, sheeley2013using, panasenco2012origins, panasenco2014apparent}. Figure \ref{fig:dome} presents two cartoons, based on 3D potential field modeling, showing how bubble shape depends on the emerging and ambient magnetic fields: the left panel reproduces a symmetric magnetic bubble with negative polarity emerging into a mean vertical positive background field (as of CH for example); the right panel represents the shape of a positive central polarity bubble emerging into a horizontal magnetic field (as in the prominence in Figure \ref{fig:SOT}). Magnetization therefore provides a natural explanation for the bubble asymmetry seen in prominences.
PSP must connect to open field regions of the bubbling magnetic carpet, modulated 
by the network of enhanced magnetic field (appearing as dark concentrations of contour lines, or nodes, in the left column of Figure \ref{fig:sources}) at the intersections of supergranules. 
They are distributed highly non-uniformly above the solar surface (at the height of 14Mm) and completely disappear at heights above 35-40 Mm, which correspond to the heights at which bubbles collapse. 
From left column of Figure \ref{fig:sources} one can see that clustering of magnetic nodes is a regular phenomenon and can also be observed along coronal hole boundaries, 
providing conditions for solar wind variability at the source.

Given these observations, we propose that observed switchback patches combine a temporal and spatial component, with the temporal component corresponding to the ``breathing'' of the magnetic carpet as observed in the bubbles seen against quiescent prominences with Hinode SOT. The typical timescale of the switchback patches (4-6 hours) is very similar with that of the magnetic bubbles and the weaker baseline field seen in patches would agree with the emergence of minor polarity flux. Also, the higher Alfv\'enicity in patches could be reconciled with interchange reconnection providing a supplementary source of Alfv\'enic fluctuations in the lower corona. The large number of individual switchbacks inside each switchback patch may be generated by interchange reconnection, which is stronger during the emergence of the bubbles when more magnetic fluxes emerge. Alternatively, the individual switchbacks may be a result of the large number of small-scale upflow plasma plumes generated at the top boundary of the bubbles due to Rayleigh-Taylor instability \citep{berger2010quiescent}. These plumes are susceptible to Kelvin-Helmholtz instability which may result in ``role-up'' of magnetic field lines, generating switchbacks. However many other details of the PSP observations, e.g. the higher proton temperature in the switchback patches, the higher baseline solar wind speed, the alpha-particle variability, are more difficult to understand and require further analysis in conjunction with observations of the corresponding source regions at the Sun. Moreover, future study of the generation frequency of the prominence bubbles is necessary. In particular, as the quiescent intervals are in general much shorter than the switchback patches (Section \ref{sec:sb_and_quiet}) as observed by PSP, we need to show that the interval between two consecutive bubbles is shorter than the bubble lifetime. At least for the event shown in Figure \ref{fig:SOT}, the interval between the collapse of the first bubble and the emergence of the second bubble is around one hour, shorter than the lifetime of the bubbles, which is roughly four hours (see Figure 1 of \citep{berger2017quiescent} that contains higher cadence snapshots of the same event).

\section{Conclusion}\label{sec:conclusion}
In this study, we have inspected PSP data from its first ten encounters to analyze switchback patches, large-scale modulations of magnetic switchbacks \citep{bale2021solar,fargette2021characteristic}. By comparing observations during three time intervals, one in encounter 1, one in encounter 10, and the other in encounter 6, we find that switchback patches may have an origin which is not purely spatial, but is also associated with some transient processes that last for several hours in the solar corona. While magnetic funnels on the supergranulation scales \citep{bale2021solar} remain a prime candidate for the patches, the observation of patches with similar time-scales during the PSP co-rotation periods point to the possibility of an intrinsically time-dependent contribution. We propose that the cycle of emerging flux, as exemplified by the bubbles appearing in contrast growing below and into quiescent prominences seen by Hinode SOT, may be a source. The spatial scales are the same as those of the supergranular network, and  the lifetime, usually 3-5 hours, are similar. Further studies should be carried out to address this conjecture, including detailed analyses of switchback patch shape (preferentially steepened edges) and their relation to the morphology of the source and the PSP orbital period, to determine how the source of the wind measured by PSP is located with respect to the coronal hole shape and boundary, as well as periods of fast wind with no clear observed patches.

\begin{acknowledgments}
This work is supported by NASA HERMES DRIVE Science Center grant No. 80NSSC20K0604 and the NASA Parker Solar Probe Observatory Scientist grant NNX15AF34G. O. Panasenco was supported by the NASA grant No. 80NSSC20K1829. The instruments of PSP were designed and developed under NASA contract NNN06AA01C. We thank the PSP SWEAP team lead by J. Kasper and FIELDS team lead by S. Bale for use of data. The level-2 magnetometer data can be found at \href{https://hpde.io/NASA/NumericalData/ParkerSolarProbe/FIELDS/MAG/Level2/RTN/FullResolution/PT0.003413S}{DOI: 10.48322/0yy0-ba92}, the level-3 SPAN proton and alpha data can be found at \href{https://hpde.io/NASA/NumericalData/ParkerSolarProbe/SWEAP/SPAN-A/Level3/ProtonPartialMoments/InstrumentFrame/PT7S}{DOI: 10.48322/ypyh-s325} and \href{https://hpde.io/NASA/NumericalData/ParkerSolarProbe/SWEAP/SPAN-A/Level3/AlphaPartialMoments/InstrumentFrame/PT14S}{DOI: 10.48322/ke19-2789}, and the level-3 SPC data can be found at \href{https://hpde.io/NASA/NumericalData/ParkerSolarProbe/SWEAP/SPC/Level3/SolarWindMomentsFits/PT0.2185S.html}{DOI: 10.48322/49we-tr31}.
\end{acknowledgments}

\bibliography{references}{}
\bibliographystyle{aasjournal}

\end{CJK*}
\end{document}